\documentclass[default]{sn-jnl}% Default
\usepackage{graphicx}%
\usepackage{multirow}%
\usepackage{tabulary}
\usepackage{amsmath,amssymb,amsfonts}%
\usepackage{amsthm}%
\usepackage{mathrsfs}%
\usepackage[title]{appendix}%
\usepackage{textcomp}%
\usepackage{manyfoot}%
\usepackage{booktabs}%
\usepackage{algorithm}%
\usepackage{algorithmicx}%
\usepackage{algpseudocode}%
\usepackage{listings}%
\usepackage[table,xcdraw]{xcolor}
\graphicspath{ {Figures/} }
\theoremstyle{thmstyleone}%
\theoremstyle{thmstyletwo}%

\theoremstyle{thmstylethree}%

\begin{document}

\title{Assessing Research Impact in Indian Conference Proceedings: Insights from Collaboration and Citations}


\author[1]{\fnm{Kiran} \sur{Sharma}}\email{kiran.sharma@bmu.edu.in}

\author*[2]{\fnm{Parul} \sur{Khurana}}\email{parul.khurana@lpu.co.in}

\affil[1]{\orgdiv{School of Engineering and Technology}, \orgname{BML Munjal University}, \orgaddress{ \street{67th Milestone, NH 48, Kapriwas}, \city{Gurugram}, \postcode{122413}, \state{Haryana}, \country{India}}}

\affil*[2]{\orgdiv{School of Computer Applications}, \orgname{Lovely Professional University}, \orgaddress{ \street{Jalandhar-Delhi, G.T. Road}, \city{Phagwara}, \postcode{144411}, \state{Punjab}, \country{India}}}


\abstract{Conferences serve as a crucial avenue for scientific communication. However, the increase in conferences and the subsequent publication of proceedings have prompted inquiries regarding the research quality being showcased at such events.  This investigation delves into the conference publications indexed by Springer's Lecture Notes in Networks and Systems Series. Among the 570 international conferences held worldwide in this series, 177 were exclusively hosted in India. These 177 conferences collectively published 11,066 papers as conference proceedings. All these publications, along with conference details, were sourced from the Scopus database. The study aims to evaluate the research impact of these conference proceedings and identify the primary contributors. The results reveal a downward trend in the average number of citations per year. The collective average citation for all publications is 1.01. Papers co-authored by Indian and international authors (5.6\%) exhibit a higher average impact of 1.44, in contrast to those authored solely by Indian authors (84.9\%), which have an average impact of 0.97. Notably, Indian-collaborated papers, among the largest contributors, predominantly originate from private colleges and universities. Only 19\% of papers exhibit collaboration with institutes of different prestige, yet their impact is considerably higher as compared to collaboration with institutes of similar prestige. This study highlights the importance of improving research quality in academic forums.}
\keywords{Conference Publications, Research Impact, Institutional Collaboration, International Participation, LNNS}

\maketitle
\section{Introduction}	
\label{IN}

Academic conferences play a vital role in the socialization of young scholars, facilitating both scientific and societal impact~\citep{cherrstrom2012making}. The participation of young scientists in conferences is an integral part of their learning journey, enabling them to broaden their knowledge and research skills, formulate networking strategies, and build relationships with fellow professionals~\citep{harrison2010unique}.

For academic scholars, maintaining high levels of research productivity is crucial for their professional advancement. In the field of computer science, conferences hold significant weight and are frequently utilized to gauge research quality. However, assessing the quality of academic research and the performance of individual researchers has long been a topic of debate~\citep{vrettas2015conferences}. While some researchers contend that relying on simplistic performance metrics for measuring research impact is inadequate and could potentially lead to manipulation of the research process, others argue that there must be some reasonable means of evaluating research performance. They posit that even an imperfect system can provide insights into research quality, aiding funding agencies and tenure committees in making more informed decisions~\citep{drott1995reexamining, liao2011improve}.

The concept of `impact' often prompts discussions regarding potential limitations of conventional approaches to measure research impact~\citep{penfield2014assessment}. Establishing a link between research and impact is inherently complex, involving social interactions~\citep{martin2011research}, and different assumptions about this process can yield varying conclusions~\citep{greenhalgh2016research}. Consequently, the assessment of the impact of conference publications, especially in comparison to journal publications, remains a subject of debate. Additionally, strategies to maximize the impact of papers published in these conferences are also a matter of ongoing discussion~\citep{hauss2021social}.

Researchers have attempted to compare the two types of publications. Rahm and Thor (2005) investigated two highly esteemed database conferences and three equally significant journals. Their findings revealed that conference papers, on an average, received higher citations compared to journal papers~\citep{rahm2005citation}. In contrast, Franceschet (2010) conducted an analysis demonstrating that while most academics publish the majority of their work in conferences, the majority of citations originate from their journal publications~\citep{franceschet2010role}. The discrepancy in conclusions between these two studies is due to the choice of data source for citations: Rahm and Thor utilized Google Scholar, while Franceschet employed the Web of Science as the selection of journals in both databases is different.

On the other side, conferences offer undeniable benefits such as rapid and regular paper publication and the opportunity for researchers to convene and engage in paper presentations and discussions with peers~\citep{birman2009viewpoint}. However, various criticisms of the conference system have been raised, especially when compared to journal publication. Some of the highlighted flaws include limited time for referees to review papers, constraints on the number of pages for publication, insufficient time for authors to refine papers after receiving reviewer comments, and the burden placed on top researchers who serve as reviewers in conference program committee~\citep{eckhaus2018improving}.

Lisee and Lariviere demonstrated that, as anticipated, conference proceedings tend to age more rapidly than cited scientific literature overall. Their findings indicate that conference proceedings have a relatively modest scientific impact, accounting for only 2\% of total citations on an average. Furthermore, their relative significance is diminishing, and they become outdated more faster than scientific literature in general~\citep{lisee2008conference}.

Another question that arises is whether there is any correlation between a conference's acceptance rate and the impact of its papers. Chen and Konstan, (2010) have suggested that the acceptance rate is one of the few variables that can be managed to enhance the impact of conferences. The authors have demonstrated a significant difference in citation counts across various acceptance rates, indicating a clear trend towards higher citations for conferences with lower acceptance rates. They have noted that conferences with low acceptance rates in computer science tend to have a greater impact than the average ACM journals~\citep{chen2010conference}.

Another inquiry raised was regarding the comparative impact of conference papers versus journal papers~\citep{freyne2010relative, gonzalez2011articles}. The distinction between conference and journal publications has often been debated in terms of article quality~\citep{kim2019author}. The authors responded that conference papers are more frequently published than journal papers, with an average of 59\% of author publications appearing in conference proceedings and 41\% in journals. Notably, high-impact scholars tend to publish significantly less frequently but more commonly in journals. 78\% of publications covering trending topics are conference papers, with 28\% being journal articles. However, journal papers receive more citations (67\%) than conference papers (33\%), indicating that journal papers have a substantially higher impact, averaging 5.41 citations per paper compared to only 0.71 citations per paper for conference papers.

Prior research has explored the impact of papers published in conferences and journals, addressing factors such as conference acceptance rates, conference venue, and the significance of conference publications in computer science~\citep{vrettas2014conferences}. In this study, we specifically investigate conference publications managed by the Springer Lecture Notes in Networks and Systems (LNNS), a Series known for its interdisciplinary approach. Our focus lies in analyzing conferences hosted in India and assessing the impact of papers published within this lecture note series. Additionally, we examine institutional collaborations and analyze the relationship between the type of collaboration and research impact. Despite these conferences being hosted in India, their internationalization prompts us to analyze international contributions and their respective research impact as well.

\subsection{Motivation}
Stimulating domains like artificial intelligence (AI), the Internet of Things (IoT), and blockchain (BC) have proven to be rapidly evolving areas of research utilized in human intelligence-based simulations in computer science~\citep{chamola2020comprehensive}. The concept of harnessing the potential of these technologies has become a topic of significant interest among researchers, driving further exploration to swiftly disseminate information to stakeholders~\citep{rabah2018convergence}.

Given that these technologies have continually provided numerous real-time solutions, including screening, anticipation, and modeling, conferences have emerged as the primary venues for examining the competitiveness of these technologies in India~\citep{khurana2023exploring}. Conference series such as Lecture Notes in Networks and Systems~\url{(https://www.springer.com/series/15179)}, Advances in Intelligent Systems and Computing~\url{(https://www.springer.com/series/11156)}, Lecture Notes in Computer Science including sub-series in Artificial Intelligence and Bioinformatics~\url{(https://www.springer.com/series/0558)}, etc.  have emerged as top contributors in Indian context~\url{(https://www.scopus.com/sources.uri)}. This prompted us to examine the papers published in the most prominent conference within the Indian context. Therefore, we opted to analyze the papers published in the Lecture Notes in Networks and Systems series.

\subsection{Research Objective}
The research objectives are:
\begin{itemize}
\item Identify the conferences within the Springer series of Lecture Notes in Networks and Systems.

\item Analyze the papers presented at conferences held in India and later published in the Springer series of Lecture Notes in Networks and Systems.

\item Investigate institutional collaboration and associated publications.

\item Explore the participation of international contributors in conferences hosted in India and their impact.

\end{itemize}

%=====================================================================================
\section{Methodology}\label{MT}
\subsection{Data Collection}\label{DC}	
The series ``Lecture Notes in Networks and Systems (LNNS)'' is a Springer series that publishes the latest developments in Networks and Systems~\url{(https://www.springer.com/series/15179)}. This is a Scopus-indexed book series. Hence, the primary source of our data is the ``Scopus database'', and data was extracted in April 2024 through an on-line interface provided by Scopus at~\url{(https://www.scopus.com/home.uri)}. As a result, a total of 570 conferences with 49,293 documents were found, which were listed in the different book volumes in this series from 2016 to 2024. The fetched information in these 570 conferences includes their name, volume, citations, country locations, themes, document titles, document types, conference dates, author names, and their affiliations primarily.

\subsection{Country Extraction}\label{CE}	
The next step was to extract the country-location information for all the fetched conferences. For 130 conferences, location was specified as ``Virtual, Online'' and for 179 conferences, cities, states, etc. were mentioned instead of countries. For all of these conferences, country location information was obtained by searching on Google~\url{(https://www.google.com/)}, conference websites, and publisher websites as well.

\subsection{Document Filtration}\label{DF}	
In this round of document filtration, we filtered conferences hosted in India. There were 177 conferences listed in this category. We picked documents with the type ``conference papers" for these 177 conferences, yielding a record of 11,066 documents spanning the years 2019 to 2024.

\subsection{Affiliation Type Extraction}\label{ATE}	
In the final step, affiliation type of authors was extracted. We visited websites of governing bodies such as the University Grants Commission of India (UGC), the Ministry of Education of India (MOE), and the All India Council for Technical Education (AICTE), along with concerned universities and colleges, to extract the affiliation type information of author affiliations listed in 11,066 documents.

For documents associated with universities, National Institute of Technology (NITs), Indian Institutes of Technology (IITs), Indian Institutes of Information Technology (IIITs), Council of Scientific \& Industrial Research (CSIRs), National Institute of Technical Teachers Training \& Research (NITTTRs), Centre for Development of Advanced Computing, India (CDACs), National Institute of Electronics and Information Technology (NIELITs), National Forensic Sciences University (NFSUs), Indian Institute of Science (IISCs), Indian Institutes of Management (IIMs), and Indian Institute of Engineering Science and Technology (IIESTs), we extracted the affiliation information through python code~\url{(https://www.python.org/)}. We merged all national importance institutes such as IITs, IIITs, NITs, etc. as ``National Institutes''.

\begin{figure}[!h]
    \centering
    \includegraphics[width=\linewidth]{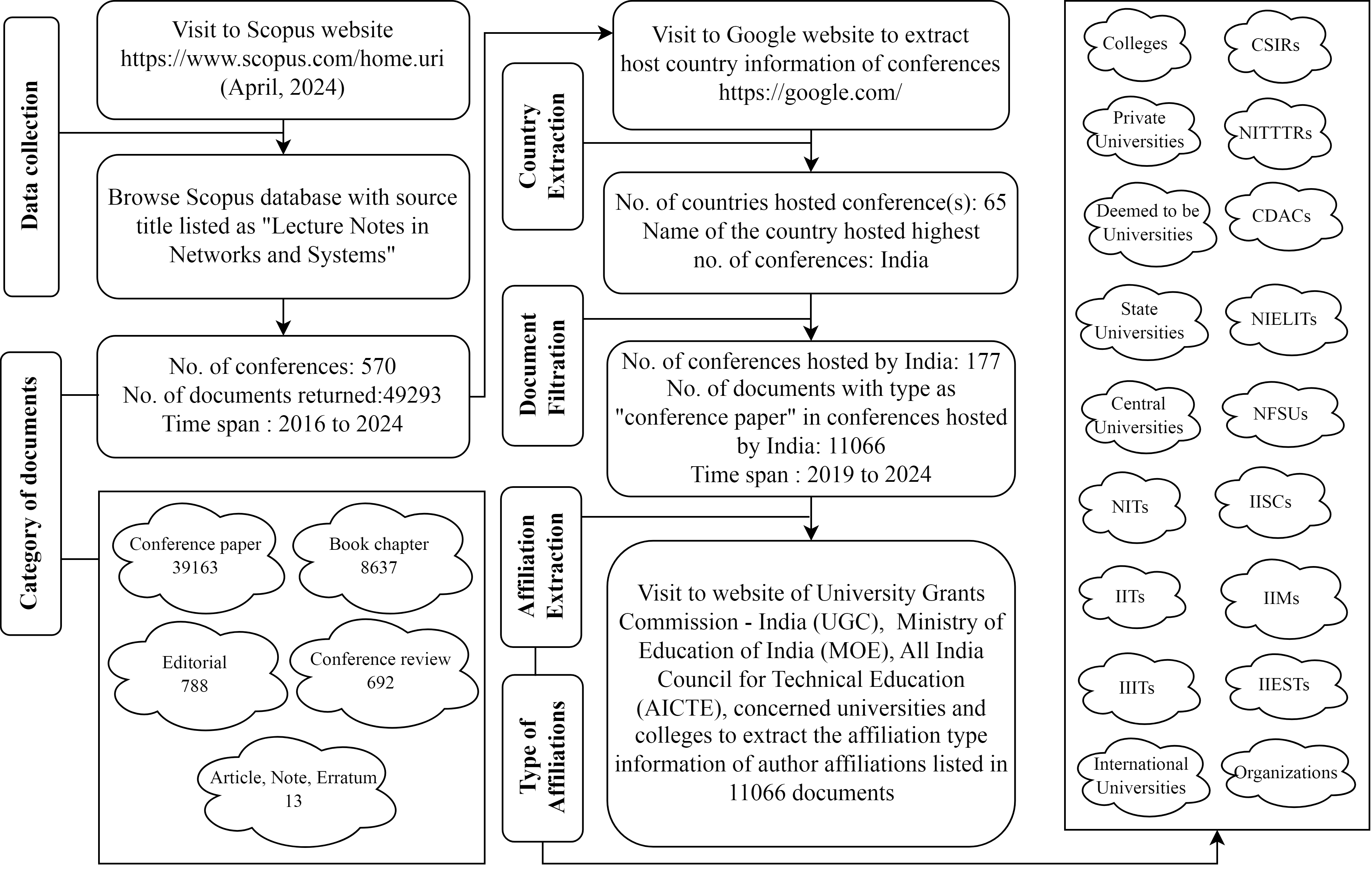}
\caption{Flowchart of data collection, filtration and feature extraction.}
\label{fig:data_flow}   
\end{figure}

But for the affiliation names listed in abbreviations, with departments, or with incomplete or partial data, we checked every record manually and listed the particular affiliations in the respective category. Figure~\ref{fig:data_flow} describes the complete process of data collection, country extraction, document filtration, and affiliation type extraction.

%%~~~~~~~~~~~~~~~~~~~~~~~~~~~~~~~~~~~~~~~~~~~~~~~~~~~~~~~~~~~~~~~~~~~~~~~~~~~~~~~~~~~~~~~~~~~~~~~~~~~~~
%%~~~~~~~~~~~~~~~~~~~~~~~~~~~~~~~~~~~~
\section{Number of Conferences and Corresponding Publications}
\subsection{Conferences Convened Worldwide}

\begin{figure}[htbp]
    \centering
\includegraphics[width=0.48\linewidth]{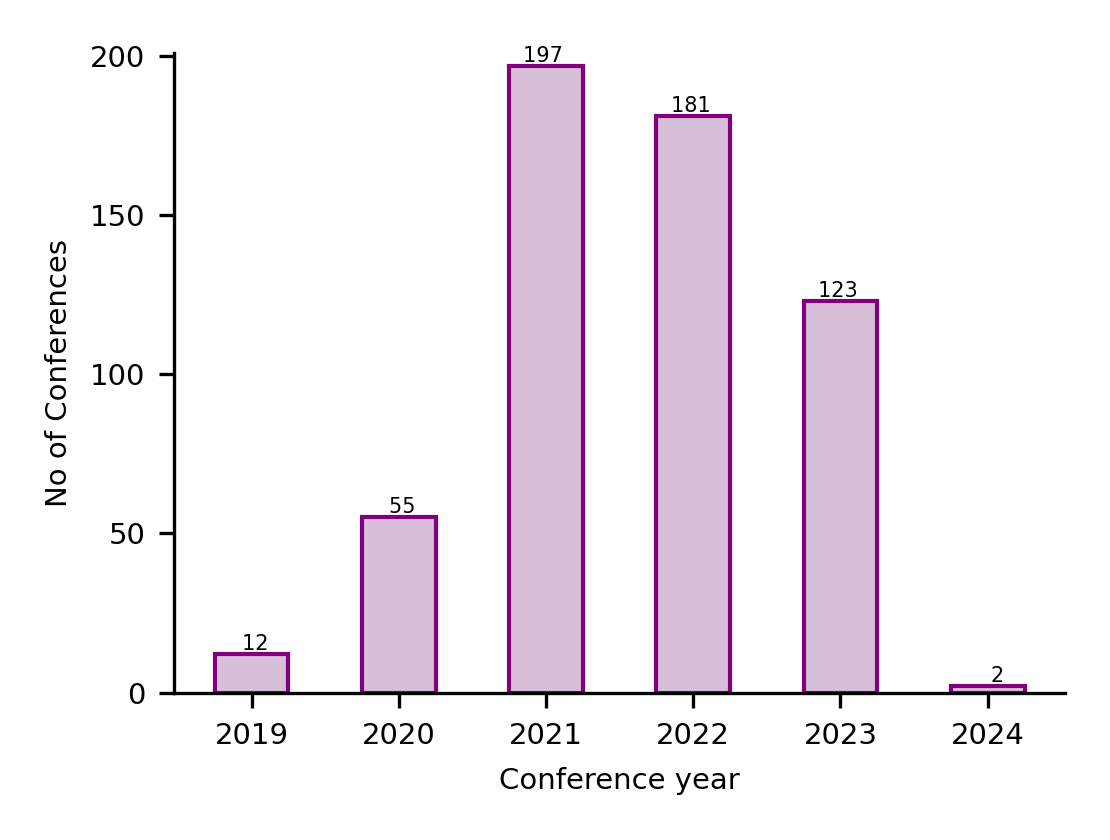}
\llap{\parbox[b]{2.5in}{(a)\\\rule{0ex}{2.0in}}}
    \includegraphics[width=0.48\linewidth]{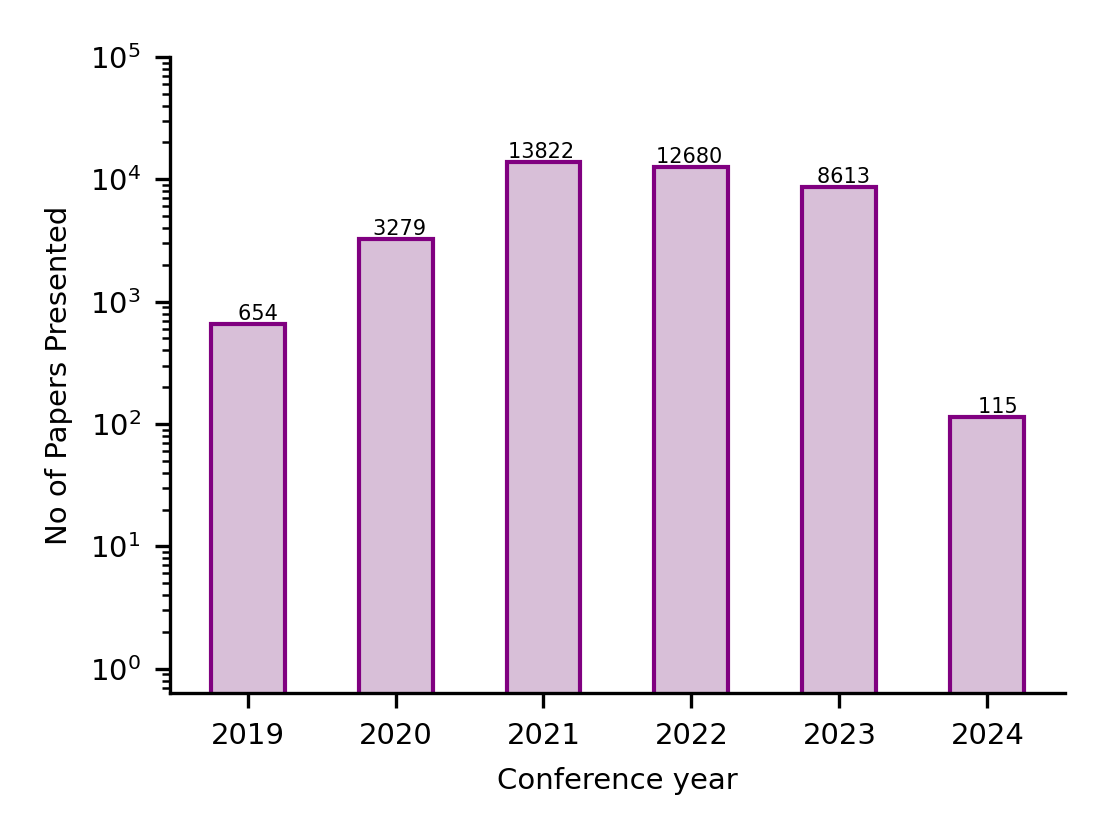}
\llap{\parbox[b]{2.5in}{(b)\\\rule{0ex}{2.0in}}}
\includegraphics[width=0.85\linewidth]{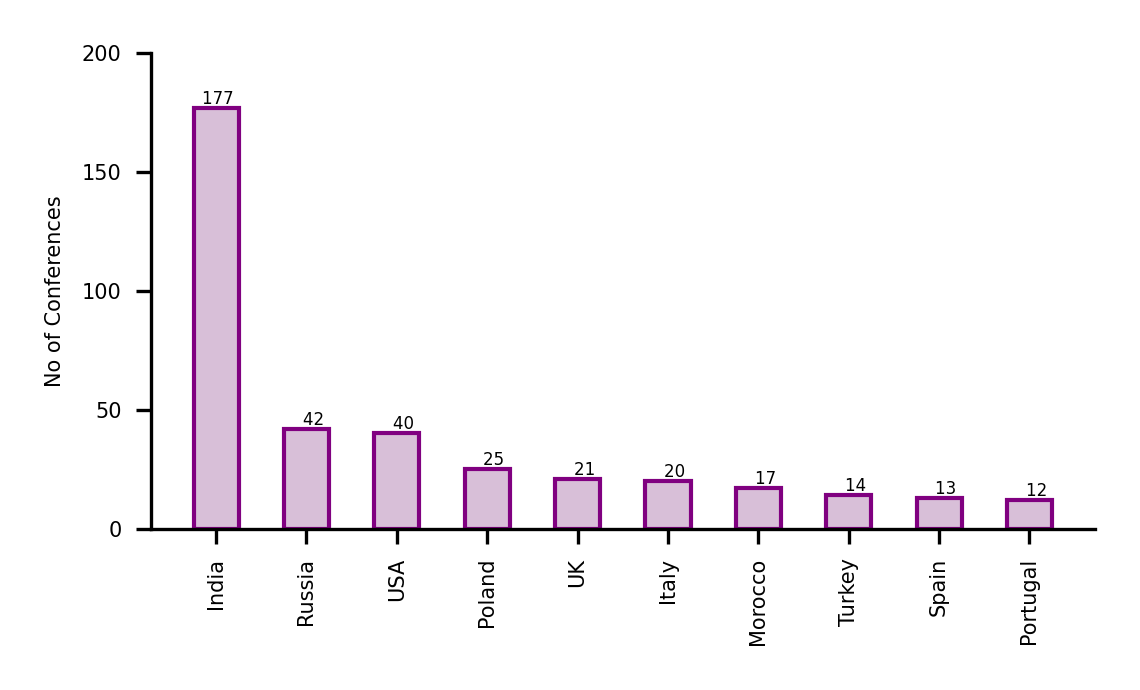}
    \llap{\parbox[b]{4.5in}{(c)\\\rule{0ex}{2.5in}}}
\caption{(a) Number of conferences held globally from 2019 to 2024. (b) Number of papers presented (later published) in all conferences held globally. (c) Top 10 countries as per number of conferences.}
\label{fig:World_Count}   
\end{figure}

Figure~\ref{fig:World_Count} illustrates the global conference landscape. In Figure~\ref{fig:World_Count}(a), the distribution of conferences from 2019 to January 2024 is depicted. Notably, the year 2021 saw a significant surge with 197 conferences, followed by 2022 with 181, and 2023 with 123. Before 2020, lecture series were predominantly published as book chapters, explaining the lower counts in those years. The uptick in conference numbers in 2021 and 2022 can be attributed to the shift to online or hybrid formats during the COVID-19 pandemic and global lockdowns, reflecting increased productivity during that period. In 2021, 13,822 papers were presented globally across 197 conferences, followed by 12,680 papers in 181 conferences in 2022, and 8,613 papers in 123 conferences in 2023. This prompts the question: who were the top contributors globally, organizing the majority of conferences, and during which periods.

Figure~\ref{fig:World_Count}(c) displays the number of conferences held in the top 10 countries. Out of a total of 570 conferences held across 65 countries, India emerged as the top contributor with 177 conferences, followed by Russia (42), USA (40), Poland (25), and others. Notably, 31\% of conferences were solely hosted by India, either in hybrid or offline formats. This significant contribution from India raises intriguing questions about the reasons behind such a rise and needs further investigation.

%%%~~~~~~~~~~~~~~~~~~~~~~~~~~~~~~~~~~~~

%%~~~~~~~~~~~~~~~~~~~~~~~~~~~~~~~~~~~~

%=========================================================================

\subsection{Conferences Hosted in India}

\begin{figure}[!h]
    \centering
    \includegraphics[width=0.48\linewidth]{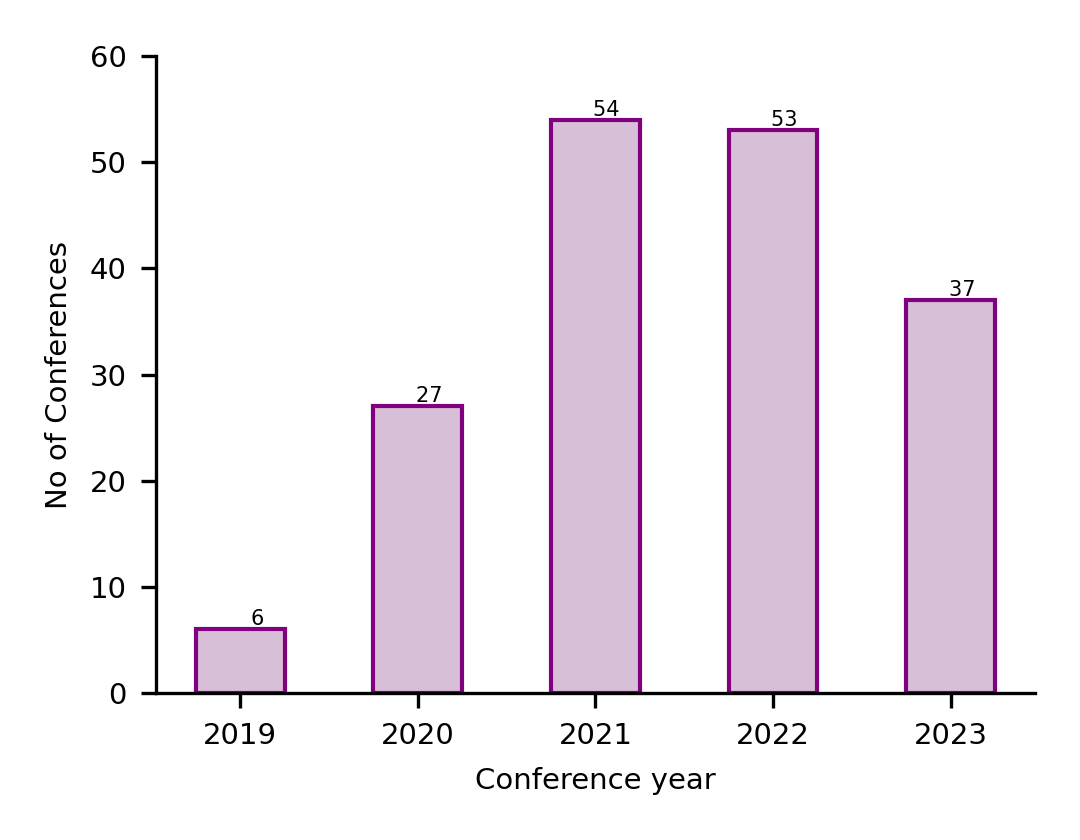}      
      \llap{\parbox[b]{2.7in}{(a)\\\rule{0ex}{1.7in}}}
    \includegraphics[width=0.48\linewidth]{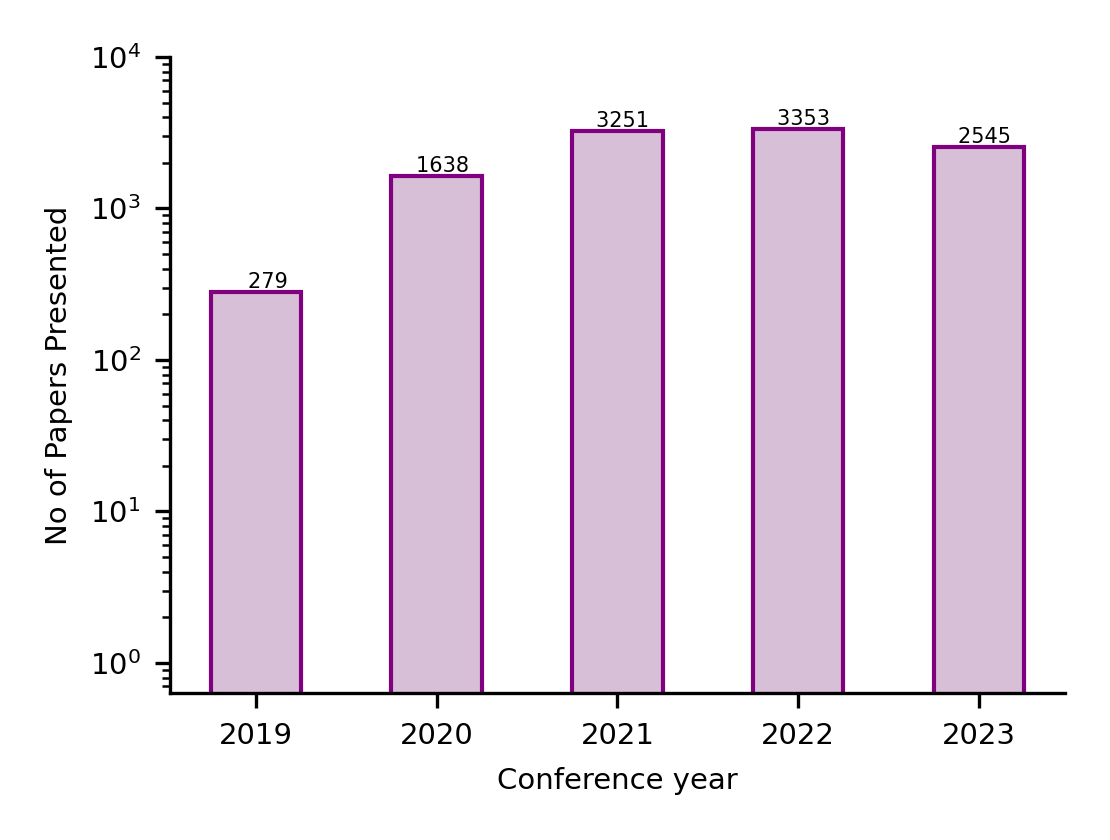}
     \llap{\parbox[b]{2.7in}{(b)\\\rule{0ex}{1.7in}}}
\caption{(a) Number of conferences held in India from 2019 to 2023. (b) Number of papers presented (later published) in all conferences held in India.}
\label{fig:India_Count}   
\end{figure}

India contributed 28.25\% of the total papers in the lecture series. In Figure~\ref{fig:India_Count}(a), the yearly count of conferences is depicted. There was a nearly equal number of conferences held in 2021 and 2022, followed by a decline in 2023. This decline could be attributed to the nature of the conferences held during this period. The majority of conferences from 2021 to 2022 were conducted online or in hybrid formats. Organizing conferences typically involves significant budget allocations, which were significantly reduced during this period, contributing to the rise in conference counts. Conversely, in Figure~\ref{fig:India_Count}(b), the number of papers contributed to these conferences is illustrated. A similar trend is observed, with almost an equal number of papers contributed in 2021 and 2022. In 2023, these contributions accounted for 23\% of the total papers. The  2021-22 time period eliminated the need for authors to travel and participate physically. This flexibility likely led to increased conference participation, as logistical and financial barriers were reduced, allowing for more frequent events.

%%%~~~~~~~~~~~~~~~~~~~~~~~~~~~~~~~~~~~~

%%~~~~~~~~~~~~~~~~~~~~~~~~~~~~~~~~~~~~
%=========================================================================
\section{Results and Discussions: Conferences Hosted in India}

\subsection{Impact of Published Papers}
\begin{figure}[!h]
    \centering
    \includegraphics[width=0.75\linewidth]{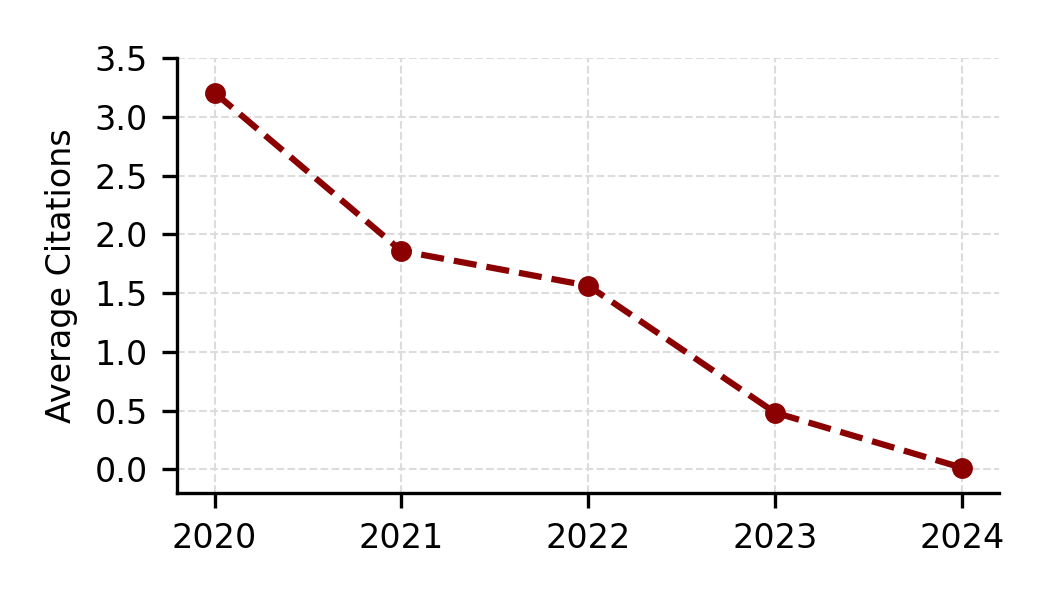}
\caption{Average citations received yearwise.}
\label{fig:India_Avg_Citations}   
\end{figure}

Over the past five years, we've observed a significant increase in paper contributions. However, the focus now shifts to the impact of these contributions, which is measured by the number of citations received by these publications. Figure~\ref{fig:India_Avg_Citations} illustrates the yearly trend of average citations for published papers. On average, the overall impact of all publications stands at 1.01.  64.5\% of papers received no citations, and 17.5\% received only one citation. Only 35.5\% publications are such which receive at least one citation. Whether that citation is an original citation or a self-citation is not known.
%%%~~~~~~~~~~~~~~~~~~~~~~~~~~~~~~~~~~~~
%%%~~~~~~~~~~~~~~~~~~~~~~~~~~~~~~~~~~~~

\subsection{Who are the Prime Contributors?}

With the overall impact of publications being notably low, it becomes imperative to examine the nature of contributions received in these conferences. Therefore, we analyzed collaboration types. Three distinct types of contributions have been identified: papers authored solely by Indian collaborators, papers authored solely by international collaborators, and papers authored collaboratively by both Indian and international authors.
The breakdown of contributions reveals that 85\% of the presented papers originate solely from Indian authors, while a mere 9.4\% are solely contributed by international authors. A smaller fraction, constituting 5.6\%, is jointly authored by both Indian and international collaborators. The average impact of publications co-authored by Indian and international authors stands higher at 1.44, surpassing those authored solely by Indian or international authors. Additionally, publications solely authored by international authors exhibit a higher impact at 1.09 compared to those authored solely by Indian authors at 0.97.

\subsection{International Participation in Conferences}

In today's landscape, nearly every conference is labelled as international. However, as observed in Table \ref{table:colab_type}, it's apparent that the proportion of international contributions is lower as compared to national ones. The number of international papers co-authored with Indian authors is notably lower than those co-authored with international counterparts. Hence, it's crucial to ascertain which countries are contributing and to what extent.
%
%%=========================================================================
% \usepackage{multirow}
% \usepackage{booktabs}
\begin{table}[]
\caption{Collaboration type and total number of papers, total citations and average citations.}
\label{table:colab_type}
\begin{tabular}{@{}lccc@{}}
\toprule
\textbf{Collaboration Type}                            & \textbf{Total Papers}     & \textbf{Total Citations}  & \textbf{Average Citations} \\ \midrule
\multicolumn{1}{|l|}{Indian and International Authors} & \multicolumn{1}{c|}{623}  & \multicolumn{1}{c|}{899}  & \multicolumn{1}{c|}{1.44}  \\ \midrule
\multicolumn{1}{|l|}{Only Indian Authors}              & \multicolumn{1}{c|}{9403} & \multicolumn{1}{c|}{9139} & \multicolumn{1}{c|}{0.97}  \\ \midrule
\multicolumn{1}{|l|}{Only International Authors}       & \multicolumn{1}{c|}{1040} & \multicolumn{1}{c|}{1139} & \multicolumn{1}{c|}{1.09}  \\ \midrule
Total                                                  & 11066                     & 11177                     & 1.01                       \\ \bottomrule
\end{tabular}
\end{table}
%=========================================================================
Table \ref{table:Top_Country} delineates the top 10 countries across various categories: overall papers, papers solely authored by international authors, and papers co-authored by Indian and international authors. A total of 95 countries have contributed, with USA leading with 161 papers, followed closely by Bangladesh (150), Iraq (142), Malaysia (99), Saudi Arabia (98), Vietnam (87), and others. Bangladesh emerges as a significant contributor with 134 papers, where authors collaborated with peers from Bangladesh or 94 other countries but not with Indian authors. Similarly, in joint publications with Indian authors, the USA takes the lead with 113 collaborations, followed by Saudi Arabia (52), Malaysia (46), Iraq (40), the United Kingdom (36), the UAE (21), and more.
%%=========================================================================
%%% List of top 10 countries based on collaboration
% \usepackage{booktabs}
\begin{table}[]
\caption{List of top 10 countries based on total international contribution,  countries with only international authors collaborated, and countries where Indian authors collaborated.}
\label{table:Top_Country}
\begin{tabular}{@{}|lc|l|lc|l|lc|l|@{}}
\toprule
\multicolumn{2}{|c|}{\textbf{\begin{tabular}[c]{@{}c@{}}Total International\\ Contributions\end{tabular}}} & \textbf{} & \multicolumn{2}{c|}{\textbf{\begin{tabular}[c]{@{}c@{}}Only International\\ Authors\end{tabular}}} & \textbf{} & \multicolumn{2}{c|}{\textbf{\begin{tabular}[c]{@{}c@{}}Authors in Collaboration\\ with India\end{tabular}}} &  \\ \midrule
\multicolumn{1}{|l|}{\textbf{Country}}                             & \textbf{Count}                        & \textbf{} & \multicolumn{1}{l|}{\textbf{Country}}                       & \textbf{Count}                       & \textbf{} & \multicolumn{1}{l|}{\textbf{Country}}                              & \textbf{Count}                         &  \\ \midrule
\multicolumn{1}{|l|}{United States}                                & 161                                   &           & \multicolumn{1}{l|}{Bangladesh}                             & 134                                  &           & \multicolumn{1}{l|}{United States}                                 & 113                                    &  \\ \midrule
\multicolumn{1}{|l|}{Bangladesh}                                   & 150                                   &           & \multicolumn{1}{l|}{Iraq}                                   & 102                                  &           & \multicolumn{1}{l|}{Saudi Arabia}                                  & 52                                     &  \\ \midrule
\multicolumn{1}{|l|}{Iraq}                                         & 142                                   &           & \multicolumn{1}{l|}{Vietnam}                                & 70                                   &           & \multicolumn{1}{l|}{Malaysia}                                      & 46                                     &  \\ \midrule
\multicolumn{1}{|l|}{Malaysia}                                     & 99                                    &           & \multicolumn{1}{l|}{Malaysia}                               & 53                                   &           & \multicolumn{1}{l|}{Iraq}                                          & 40                                     &  \\ \midrule
\multicolumn{1}{|l|}{Saudi Arabia}                                 & 98                                    &           & \multicolumn{1}{l|}{Indonesia}                              & 50                                   &           & \multicolumn{1}{l|}{United Kingdom}                                & 36                                     &  \\ \midrule
\multicolumn{1}{|l|}{Vietnam}                                      & 87                                    &           & \multicolumn{1}{l|}{United States}                          & 48                                   &           & \multicolumn{1}{l|}{United Arab Emirates}                          & 21                                     &  \\ \midrule
\multicolumn{1}{|l|}{Indonesia}                                    & 57                                    &           & \multicolumn{1}{l|}{South Africa}                           & 48                                   &           & \multicolumn{1}{l|}{Ethiopia}                                      & 20                                     &  \\ \midrule
\multicolumn{1}{|l|}{United Kingdom}                               & 57                                    &           & \multicolumn{1}{l|}{Saudi Arabia}                           & 46                                   &           & \multicolumn{1}{l|}{Australia}                                     & 19                                     &  \\ \midrule
\multicolumn{1}{|l|}{Japan}                                        & 53                                    &           & \multicolumn{1}{l|}{Japan}                                  & 40                                   &           & \multicolumn{1}{l|}{South Korea}                                   & 18                                     &  \\ \midrule
\multicolumn{1}{|l|}{United Arab Emirates}                         & 52                                    &           & \multicolumn{1}{l|}{Egypt}                                  & 40                                   &           & \multicolumn{1}{l|}{Vietnam}                                       & 17                                     &  \\ \bottomrule
\end{tabular}
\end{table}
%%=========================================================================
%=========================================================================
\subsection{Impact of Institutional Collaboration}
Papers published within the same type of institution refer to those where all the authors share affiliations with institutions of similar kinds, such as colleges collaborate with colleges, and so on. Conversely, papers published across different types of institutions involve authors affiliated with diverse institutional backgrounds, such as collaborations between authors from private and public universities. The data illustrates that 81\% of published papers predominantly involve authors from the same institute or those affiliated with similar kinds of institutes, while 19\% of papers result from collaborations among authors from different types of institutions. Notably, a significant portion of contributions comes from colleges (31.2\%), followed by private universities (17.4\%), deemed universities (16.6\%), international universities (12.3\%), state universities (10.1\%), and central universities (2.6\%). National Institutes encompass all institutes of national importance (8.1\%) such as IITs, IIITs, NITs, CSIR, etc., while a minor fraction of contributions originates from organizations (1.6\%). This analysis underscores the tendency for conference participants to primarily collaborate within their own institutions or institutions of similar prestige. Table~\ref{table:colab_type1} details the count of papers published within each category of the same type.
%=========================================================================
%*****************************************************************************
\begin{table}[]
\caption{Collaboration with institutes of same prestige: Total number of paper, average citation and count of single and multi-affiliated papers.}
\label{table:colab_type1}
\begin{tabular}{@{}lccccc@{}}
\toprule
\multirow{2}{*}{\textbf{Collaboration Type}}                              & \multicolumn{4}{c}{\textbf{No. of Publications}}                                                                                                                                                                                                                       & \multicolumn{1}{l}{\multirow{2}{*}{\textbf{\begin{tabular}[c]{@{}l@{}}Average\\ Citations\end{tabular}}}} \\ \cmidrule(lr){2-5}
                                                                          & \multicolumn{1}{c|}{\textbf{Total}} & \multicolumn{1}{c|}{\textbf{In \%}} & \multicolumn{1}{c|}{\textbf{\begin{tabular}[c]{@{}c@{}}Single\\ Affiliation\end{tabular}}} & \multicolumn{1}{c|}{\textbf{\begin{tabular}[c]{@{}c@{}}Multiple\\ Affiliations\end{tabular}}} & \multicolumn{1}{l}{}                                                                                      \\ \midrule
\multicolumn{1}{|l|}{College - College}                                   & \multicolumn{1}{c|}{2879}           & \multicolumn{1}{c|}{32.27}          & \multicolumn{1}{c|}{1828}                                                                  & \multicolumn{1}{c|}{1051}                                                                     & \multicolumn{1}{c|}{0.81}                                                                                 \\ \midrule
\multicolumn{1}{|l|}{Private University - Private University}             & \multicolumn{1}{c|}{1521}           & \multicolumn{1}{c|}{17.05}          & \multicolumn{1}{c|}{1094}                                                                  & \multicolumn{1}{c|}{427}                                                                      & \multicolumn{1}{c|}{1.14}                                                                                 \\ \midrule
\multicolumn{1}{|l|}{Deemed to be University - Deemed to be University}   & \multicolumn{1}{c|}{1458}           & \multicolumn{1}{c|}{16.34}          & \multicolumn{1}{c|}{1121}                                                                  & \multicolumn{1}{c|}{337}                                                                      & \multicolumn{1}{c|}{1.17}                                                                                 \\ \midrule
\multicolumn{1}{|l|}{International University - International University} & \multicolumn{1}{c|}{1073}           & \multicolumn{1}{c|}{12.03}          & \multicolumn{1}{c|}{517}                                                                   & \multicolumn{1}{c|}{556}                                                                      & \multicolumn{1}{c|}{0.94}                                                                                 \\ \midrule
\multicolumn{1}{|l|}{State University - State University}                 & \multicolumn{1}{c|}{884}            & \multicolumn{1}{c|}{9.91}           & \multicolumn{1}{c|}{692}                                                                   & \multicolumn{1}{c|}{192}                                                                      & \multicolumn{1}{c|}{0.75}                                                                                 \\ \midrule
\multicolumn{1}{|l|}{National Institutes - National Institutes}           & \multicolumn{1}{c|}{772}            & \multicolumn{1}{c|}{8.65}           & \multicolumn{1}{c|}{585}                                                                   & \multicolumn{1}{c|}{187}                                                                      & \multicolumn{1}{c|}{0.92}                                                                                 \\ \midrule
\multicolumn{1}{|l|}{Central University - Central University}             & \multicolumn{1}{c|}{243}            & \multicolumn{1}{c|}{2.72}           & \multicolumn{1}{c|}{170}                                                                   & \multicolumn{1}{c|}{73}                                                                       & \multicolumn{1}{c|}{1.01}                                                                                 \\ \midrule
\multicolumn{1}{|l|}{Organization - Organization}                         & \multicolumn{1}{c|}{91}             & \multicolumn{1}{c|}{1.02}           & \multicolumn{1}{c|}{75}                                                                    & \multicolumn{1}{c|}{16}                                                                       & \multicolumn{1}{c|}{0.21}                                                                                 \\ \midrule
\textbf{Total}                                                            & \textbf{8921}                       & \textbf{-}                          & \textbf{6082}                                                                              & \textbf{2839}                                                                                 & \textbf{-}                                                                                                \\ \bottomrule
\end{tabular}
\end{table}
%=========================================================================

Papers authored exclusively by Indian authors exhibit a higher volume of publications but a lower impact compared to papers authored solely by international authors or those co-authored by Indian and international authors. The majority of contributions originate from papers affiliated with colleges (32.3\%), with an average citation count of 0.81. Conversely, private universities, with 17.05\% papers, demonstrate a higher average citation count (1.14) compared to college publications. Deemed universities represent 16.3\% of publications with the highest average citations among all (1.17). Central university contributions are relatively low at 2.7\%, yet they claim a better average citation count of 1.01 (see Table \ref{table:colab_type1}). Furthermore, 68.2\% of contributions stem from collaborations among the institutes of same prestige.

Similarly, when scrutinizing publications affiliated with institutes of varying prestige, it becomes apparent that collaborations involving international authors and publicly funded universities yield higher impacts. For instance, collaborations between International Universities and State Universities account for 1.68\% of publications with an average citation count of 2.72, while collaborations between International Universities and Private Universities (7.06\%, 1.93) also demonstrate notable impacts (see Table~\ref{table:colb_type2}). Only 19\% of papers exhibit collaboration with institutes of different prestige, yet their impact is considerably higher compared to collaborations with institutes of similar prestige.

% \usepackage{booktabs}
\begin{table}[]
\caption{Collaboration with institutes of different prestige: Total number of paper (NP), and average citation (AC).}
\label{table:colb_type2}
\begin{tabular}{@{}lccclccc@{}}
\toprule
\textbf{Collaboration Type}                                                                                        & \textbf{NP}              & \textbf{In \%}             & \textbf{AC}               & \textbf{Collaboration Type}                                                                                   & \textbf{NP}             & \textbf{In \%}            & \textbf{AC}               \\ \midrule
\multicolumn{1}{|l|}{\begin{tabular}[c]{@{}l@{}}College -\\ Deemed to be University\end{tabular}}                  & \multicolumn{1}{c|}{430} & \multicolumn{1}{c|}{15.41} & \multicolumn{1}{c|}{0.87} & \multicolumn{1}{l|}{\begin{tabular}[c]{@{}l@{}}Private University -\\ Organization\end{tabular}}              & \multicolumn{1}{c|}{44} & \multicolumn{1}{c|}{1.58} & \multicolumn{1}{c|}{1.11} \\ \midrule
\multicolumn{1}{|l|}{\begin{tabular}[c]{@{}l@{}}College -\\ Private University\end{tabular}}                       & \multicolumn{1}{c|}{382} & \multicolumn{1}{c|}{13.69} & \multicolumn{1}{c|}{1.87} & \multicolumn{1}{l|}{\begin{tabular}[c]{@{}l@{}}International University -\\ National Institutes\end{tabular}} & \multicolumn{1}{c|}{42} & \multicolumn{1}{c|}{1.5}  & \multicolumn{1}{c|}{1.39} \\ \midrule
\multicolumn{1}{|l|}{\begin{tabular}[c]{@{}l@{}}International University -\\ College\end{tabular}}                 & \multicolumn{1}{c|}{323} & \multicolumn{1}{c|}{11.57} & \multicolumn{1}{c|}{1.67} & \multicolumn{1}{l|}{\begin{tabular}[c]{@{}l@{}}National Institutes -\\ State University\end{tabular}}         & \multicolumn{1}{c|}{30} & \multicolumn{1}{c|}{1.07} & \multicolumn{1}{c|}{1.83} \\ \midrule
\multicolumn{1}{|l|}{\begin{tabular}[c]{@{}l@{}}College -\\ State University\end{tabular}}                         & \multicolumn{1}{c|}{265} & \multicolumn{1}{c|}{9.49}  & \multicolumn{1}{c|}{1.3}  & \multicolumn{1}{l|}{\begin{tabular}[c]{@{}l@{}}Deemed to be University -\\ Organization\end{tabular}}         & \multicolumn{1}{c|}{30} & \multicolumn{1}{c|}{1.07} & \multicolumn{1}{c|}{0.63} \\ \midrule
\multicolumn{1}{|l|}{\begin{tabular}[c]{@{}l@{}}Deemed to be University -\\ Private University\end{tabular}}       & \multicolumn{1}{c|}{215} & \multicolumn{1}{c|}{7.7}   & \multicolumn{1}{c|}{1.2}  & \multicolumn{1}{l|}{\begin{tabular}[c]{@{}l@{}}College -\\ Organization\end{tabular}}                         & \multicolumn{1}{c|}{27} & \multicolumn{1}{c|}{0.97} & \multicolumn{1}{c|}{0.77} \\ \midrule
\multicolumn{1}{|l|}{\begin{tabular}[c]{@{}l@{}}International University -\\ Private University\end{tabular}}      & \multicolumn{1}{c|}{197} & \multicolumn{1}{c|}{7.06}  & \multicolumn{1}{c|}{1.93} & \multicolumn{1}{l|}{\begin{tabular}[c]{@{}l@{}}State University -\\ Central University\end{tabular}}          & \multicolumn{1}{c|}{24} & \multicolumn{1}{c|}{0.86} & \multicolumn{1}{c|}{0.5}  \\ \midrule
\multicolumn{1}{|l|}{\begin{tabular}[c]{@{}l@{}}International University -\\ Deemed to be University\end{tabular}} & \multicolumn{1}{c|}{137} & \multicolumn{1}{c|}{4.91}  & \multicolumn{1}{c|}{1.22} & \multicolumn{1}{l|}{\begin{tabular}[c]{@{}l@{}}State University -\\ Organization\end{tabular}}                & \multicolumn{1}{c|}{22} & \multicolumn{1}{c|}{0.79} & \multicolumn{1}{c|}{1.04} \\ \midrule
\multicolumn{1}{|l|}{\begin{tabular}[c]{@{}l@{}}College -\\ National Institutes\end{tabular}}                      & \multicolumn{1}{c|}{101} & \multicolumn{1}{c|}{3.62}  & \multicolumn{1}{c|}{0.69} & \multicolumn{1}{l|}{\begin{tabular}[c]{@{}l@{}}International University -\\ Central University\end{tabular}}  & \multicolumn{1}{c|}{21} & \multicolumn{1}{c|}{0.75} & \multicolumn{1}{c|}{1.28} \\ \midrule
\multicolumn{1}{|l|}{\begin{tabular}[c]{@{}l@{}}Private University -\\ State University\end{tabular}}              & \multicolumn{1}{c|}{98}  & \multicolumn{1}{c|}{3.51}  & \multicolumn{1}{c|}{1.11} & \multicolumn{1}{l|}{\begin{tabular}[c]{@{}l@{}}Private University -\\ Central University\end{tabular}}        & \multicolumn{1}{c|}{19} & \multicolumn{1}{c|}{0.68} & \multicolumn{1}{c|}{0.57} \\ \midrule
\multicolumn{1}{|l|}{\begin{tabular}[c]{@{}l@{}}National Institutes -\\ Private University\end{tabular}}           & \multicolumn{1}{c|}{75}  & \multicolumn{1}{c|}{2.69}  & \multicolumn{1}{c|}{0.74} & \multicolumn{1}{l|}{\begin{tabular}[c]{@{}l@{}}National Institutes -\\ Organization\end{tabular}}             & \multicolumn{1}{c|}{14} & \multicolumn{1}{c|}{0.5}  & \multicolumn{1}{c|}{0.35} \\ \midrule
\multicolumn{1}{|l|}{\begin{tabular}[c]{@{}l@{}}Deemed to be University - \\ National Institutes\end{tabular}}     & \multicolumn{1}{c|}{73}  & \multicolumn{1}{c|}{2.62}  & \multicolumn{1}{c|}{0.91} & \multicolumn{1}{l|}{\begin{tabular}[c]{@{}l@{}}National Institutes -\\ Central University\end{tabular}}       & \multicolumn{1}{c|}{9}  & \multicolumn{1}{c|}{0.32} & \multicolumn{1}{c|}{0.66} \\ \midrule
\multicolumn{1}{|l|}{\begin{tabular}[c]{@{}l@{}}Deemed to be University - \\ State University\end{tabular}}        & \multicolumn{1}{c|}{65}  & \multicolumn{1}{c|}{2.33}  & \multicolumn{1}{c|}{0.75} & \multicolumn{1}{l|}{\begin{tabular}[c]{@{}l@{}}Deemed to be University -\\ Central University\end{tabular}}   & \multicolumn{1}{c|}{9}  & \multicolumn{1}{c|}{0.32} & \multicolumn{1}{c|}{0.22} \\ \midrule
\multicolumn{1}{|l|}{\begin{tabular}[c]{@{}l@{}}College -\\ Central University\end{tabular}}                       & \multicolumn{1}{c|}{50}  & \multicolumn{1}{c|}{1.79}  & \multicolumn{1}{c|}{0.4}  & \multicolumn{1}{l|}{\begin{tabular}[c]{@{}l@{}}Central University -\\ Organization\end{tabular}}              & \multicolumn{1}{c|}{2}  & \multicolumn{1}{c|}{0.07} & \multicolumn{1}{c|}{0.5}  \\ \midrule
\begin{tabular}[c]{@{}l@{}}International University -\\ State University\end{tabular}                              & 47                       & 1.68                       & 2.72                      & -                                                                                                             & -                       & -                         & -                         \\ \bottomrule
\end{tabular}
\end{table}
%=========================================================================
\subsection{Top 10 Conferences in the Series}
Table \ref{table:colb_type3} presents the top 10 conferences in this lecture note series held in India based on the total number of papers, along with their corresponding total citations received, average citations, and conference locations. The average impact reflects the influence of the papers published in these conferences. Despite all the top 10 conferences generating more than 100 papers each, the impact analysis is negligible. The majority of these conferences in the top 10 were held in the year 2022. Only the conference held in the year 2021 shows a slightly higher impact with an average citation of 1.56 compared to the other conferences.
%~~~~~~~~~~~~~~~~~~~~~~~~~~~~~~~~~~~~~~~~~~~~~~~~~~~~~~~~~

\begin{table}[]
\caption{Top 10 conferences as per number of papers published held in India with location, total citations and average citations.}
\label{table:colb_type3}
\begin{tabular}{@{}lcccc@{}}
\toprule
\textbf{Conference Name}                                                                                                                                                       & \textbf{Location}              & \textbf{\#Papers}        & \textbf{\begin{tabular}[c]{@{}c@{}}Total\\ Citations\end{tabular}} & \textbf{\begin{tabular}[c]{@{}c@{}}Average\\ Citations\end{tabular}} \\ \midrule
\multicolumn{1}{|l|}{\begin{tabular}[c]{@{}l@{}}6th International Conference on Innovative\\ Computing and Communication, ICICC 2023\end{tabular}}                             & \multicolumn{1}{c|}{New Delhi} & \multicolumn{1}{c|}{197} & \multicolumn{1}{c|}{49}                                            & \multicolumn{1}{c|}{0.25}                                            \\ \midrule
\multicolumn{1}{|l|}{\begin{tabular}[c]{@{}l@{}}5th International Conference on Innovative\\ Computing and Communication, ICICC 2022\end{tabular}}                             & \multicolumn{1}{c|}{New Delhi} & \multicolumn{1}{c|}{195} & \multicolumn{1}{c|}{179}                                           & \multicolumn{1}{c|}{0.92}                                            \\ \midrule
\multicolumn{1}{|l|}{\begin{tabular}[c]{@{}l@{}}8th International Conference on ICT for\\ Sustainable Development, ICT4SD 2023\end{tabular}}                                   & \multicolumn{1}{c|}{Goa}       & \multicolumn{1}{c|}{165} & \multicolumn{1}{c|}{9}                                             & \multicolumn{1}{c|}{0.05}                                            \\ \midrule
\multicolumn{1}{|l|}{\begin{tabular}[c]{@{}l@{}}19th International Conference on Humanizing\\ Work and Work Environment, HWWE 2021\end{tabular}}                               & \multicolumn{1}{c|}{Guwahati}  & \multicolumn{1}{c|}{154} & \multicolumn{1}{c|}{60}                                            & \multicolumn{1}{c|}{0.38}                                            \\ \midrule
\multicolumn{1}{|l|}{\begin{tabular}[c]{@{}l@{}}3rd International Conference on Emerging Technologies\\ in Data Mining and Information Security, IEMIS 2022\end{tabular}}      & \multicolumn{1}{c|}{Kolkata}   & \multicolumn{1}{c|}{146} & \multicolumn{1}{c|}{69}                                            & \multicolumn{1}{c|}{0.47}                                            \\ \midrule
\multicolumn{1}{|l|}{\begin{tabular}[c]{@{}l@{}}7th International Conference on Smart Trends in\\ Computing and Communications, SmartCom 2023\end{tabular}}                    & \multicolumn{1}{c|}{Jaipur}    & \multicolumn{1}{c|}{143} & \multicolumn{1}{c|}{25}                                            & \multicolumn{1}{c|}{0.17}                                            \\ \midrule
\multicolumn{1}{|l|}{\begin{tabular}[c]{@{}l@{}}International Conference on Advances in Data-driven\\ Computing and Intelligent Systems, ADCIS 2022\end{tabular}}              & \multicolumn{1}{c|}{Goa}       & \multicolumn{1}{c|}{132} & \multicolumn{1}{c|}{8}                                             & \multicolumn{1}{c|}{0.06}                                            \\ \midrule
\multicolumn{1}{|l|}{\begin{tabular}[c]{@{}l@{}}2nd International Conference on Data Science\\ and Applications, ICDSA 2021\end{tabular}}                                      & \multicolumn{1}{c|}{Online}    & \multicolumn{1}{c|}{125} & \multicolumn{1}{c|}{195}                                           & \multicolumn{1}{c|}{1.56}                                            \\ \midrule
\multicolumn{1}{|l|}{\begin{tabular}[c]{@{}l@{}}7th International Conference on Information and\\ Communication Technology for Intelligent Systems,\\ ICTIS 2023\end{tabular}} & \multicolumn{1}{c|}{Ahmedabad} & \multicolumn{1}{c|}{115} & \multicolumn{1}{c|}{10}                                            & \multicolumn{1}{c|}{0.08}                                            \\ \midrule
\multicolumn{1}{|l|}{\begin{tabular}[c]{@{}l@{}}4th International Conference on Communication\\ and Intelligent Systems, ICCIS 2022\end{tabular}}                              & \multicolumn{1}{c|}{New Delhi} & \multicolumn{1}{c|}{107} & \multicolumn{1}{c|}{160}                                           & \multicolumn{1}{c|}{0.15}                                            \\ \bottomrule
\end{tabular}
\end{table}

%=========================================================================

\section{Discussion and Conclusion}

This study examined the impact of conferences hosted in India and subsequent publications in the LNNS. It assessed the impact of the published papers, measured by their citations, as well as institutional collaboration and international contributions. The proliferation of conferences in India may be attributed to individual academic incentives, where Scopus-indexed papers are often favored for annual evaluations, enhancing academic reputations. However, amidst this emphasis on quantity, questions arise regarding the impact of such publications. Are we prioritizing quantity over impactful research? For researchers, the societal impact and utilization of their work are paramount, yet these considerations appear to be overshadowed in this context.

The present research revealed that the conference-published papers received notably low citation counts. Upon comparison with the collaborative nature of the papers, it is evident that papers authored solely by Indian researchers has less impact compared to those co-authored with international counterparts. Additionally, it is observed that the primary contributions in these conferences predominantly originated from colleges and private universities. Furthermore, the majority of collaborations within these papers occurred within the same institution or among institutions of similar prestige.

There could be various other factors contributing to the lack of impact of these publications. One potential reason for their limited impact could be their lack of open access. Antelman (2004) demonstrated that open-access articles tend to have a greater research impact compared to those behind paywalls. Another factor might be the prestige of the conference and venue\citep{antelman2004open}. Chen and Konstan, (2010) noted that papers published in highly selective computer science conferences receive more citations on average than those in transactions and journals. Additionally, the quality of papers published in conferences could be affected by page length constraints, limiting researchers from conducting more extensive research~\citep{loizides2017evaluating}. Furthermore, conferences indexed by Scopus may only accept citations from papers published in journals also indexed by Scopus. Given that Scopus's journal coverage is narrower than Google Scholar's, papers may receive fewer citations, thus capturing lower impact.

On the contrary, Google Scholar excels in covering conference proceedings compared to Scopus and Web of Science~\citep{meho2007impact}. However, Google Scholar has its drawbacks, including the potential for fake publications to be indexed and the indexing of multiple versions of the same document as distinct~\citep{harzing2014longitudinal}. Moreover, conference papers may lack the level of innovation found in journal articles, prompting to Moshe Vardi's question in a May 2009 Communications editor's letter, whether we are heading in the wrong direction concerning publication practices~\citep{vardi2009conferences}.

The analysis presented in this research work has the following limitations:

\begin{itemize}
\item Due to the limited data availability, we could not take the themes/tracks of the conferences or their publications.

\item This study analyzes only one conference series, such as LNNS, whereas the study may explore the other conference series as well.

\item This study analyzes the conference series in the Indian context only, further studies may explore the international context as well.

\end{itemize}
%===============================================================================================================
\section*{Data Availability Statement}
The data can be provided upon request from the corresponding author.
\section*{Conflict of Interest}
The author declares no conflict of interest.
\bibliographystyle{unsrtnat}
\bibliography{custom}
\end{document}